\documentclass[aps,prl,reprint,superscriptaddress,showpacs]{revtex4-1}
\usepackage{graphicx}
\usepackage{amsmath}
\usepackage{bm}
\usepackage{dcolumn}
\usepackage{amsfonts}
\usepackage{amssymb}
\usepackage[breaklinks=true]{hyperref}
\hypersetup{colorlinks=true, citecolor=blue, filecolor=blue, linkcolor=blue, urlcolor=blue}

\begin{document}

\title{Influence of yttrium iron garnet thickness and heater opacity on the nonlocal transport of electrically and thermally excited magnons}
\author{Juan Shan}
\email[]{j.shan@rug.nl}
\author{Ludo J. Cornelissen}
\author{Nynke Vlietstra}
\affiliation{Physics of Nanodevices, Zernike Institute for Advanced Materials, University of Groningen, Nijenborgh 4, 9747 AG Groningen, The Netherlands}
\author{Jamal Ben Youssef}
\affiliation{Universit\'{e} de Bretagne Occidentale, Laboratoire de Magn\'{e}tisme de Bretagne CNRS, 6 Avenue Le Gorgeu, 29285 Brest, France}
\author{Timo Kuschel}
\affiliation{Physics of Nanodevices, Zernike Institute for Advanced Materials, University of Groningen, Nijenborgh 4, 9747 AG Groningen, The Netherlands}
\author{Rembert A. Duine}
\affiliation{Institute for Theoretical Physics and Center for Extreme Matter and Emergent Phenomena, Utrecht University, Leuvenlaan 4, 3584 CE Utrecht, The Netherlands}
\affiliation{Department of Applied Physics, Eindhoven University of Technology,	PO Box 513, 5600 MB Eindhoven, The Netherlands}
\author{Bart J. van Wees}
\affiliation{Physics of Nanodevices, Zernike Institute for Advanced Materials, University of Groningen, Nijenborgh 4, 9747 AG Groningen, The Netherlands}

\date{\today}

\begin{abstract}
We studied the nonlocal transport behavior of both electrically and thermally excited magnons in yttrium iron garnet (YIG) as a function of its thickness. For electrically injected magnons, the nonlocal signals decrease monotonically as the YIG thickness increases. For the nonlocal behavior of the thermally generated magnons, or the nonlocal spin Seebeck effect (SSE), we observed a sign reversal which occurs at a certain heater-detector distance, and it is influenced by both the opacity of the YIG/heater interface and the YIG thickness. Our nonlocal SSE results can be qualitatively explained by the bulk-driven SSE mechanism together with the magnon diffusion model. Using a two-dimensional finite element model (2D-FEM), we estimated the bulk spin Seebeck coefficient of YIG at room temperature. The quantitative disagreement between the experimental and modeled results indicates more complex processes going on in addition to magnon diffusion and relaxation, especially close to the contacts.
\end{abstract}
\maketitle

\section{I.~~~ Introduction}

Magnons, the quanta of spin waves, are collective excitations of electron spin angular momentum in magnetically ordered materials. Recently, magnons entered the field of spintronics \cite{zutic_spintronics:_2004} as novel spin information carriers, opening the field of magnon spintronics \cite{chumak_magnon_2015}. Just as the study of spin-polarized electric currents, the excitation, transmission and detection of magnons are of central interest to this field. 

Though magnons exist in magnetic materials at any finite temperature below the Curie temperature $T_c$, following the Bose-Einstein distribution with a zero chemical potential, only the magnons in excess of equilibrium, i.e., the non-equilibrium magnons, can be manipulated and are relevant for spin information encoding and transmission. Non-equilibrium magnons can be excited either coherently or incoherently. Coherent precession of the magnetic moments can be generated by, for instance, ferromagnetic resonance (FMR) \cite{kittel_introduction_2004} or spin transfer torque (STT) \cite{kajiwara_transmission_2010,madami_direct_2011,demidov_magnetic_2012,collet_generation_2016}. In the frequency spectrum, these excited magnons form a narrow peak, typically in the GHz range. 

The alternative incoherent generation of magnons is attractive in that it does not require an external microwave field or a large threshold electric current density, though the frequencies
of the excited magnons cannot be well controlled and are spread out in a broad spectrum. One prominent example is the spin Seebeck effect (SSE) \cite{uchida_observation_2008,uchida_spin_2010}, the excitation of magnons by a thermal gradient applied to the magnetic material. When the magnon current flows into a neighboring metal with strong spin-orbit coupling, such as platinum (Pt), a charge current is induced as a result of the inverse spin Hall effect (ISHE). Different theories \cite{xiao_theory_2010,adachi_linear-response_2011,adachi_theory_2013,hoffman_landau-lifshitz_2013,schreier_magnon_2013,rezende_magnon_2014,duine_spintronics_2015} were proposed to explain the mechanism of the thermal excitation of the magnons; meanwhile, experimental results \cite{jaworski_spin-seebeck_2011,uchida_quantitative_2014,kikkawa_critical_2015,jin_effect_2015,vlietstra_simultaneous_2014,kehlberger_length_2015,guo_origin_2015,meier_longitudinal_2015} have revealed its complex nature. In particular, the yttrium iron garnet (YIG) thickness-dependent study \cite{kehlberger_length_2015} indicates the bulk-nature of the SSE, and shows a finite magnon diffusion length $\lambda_m$ with an upper limit of 1 $\mu$m for the YIG grown by liquid phase epitaxy (LPE) at room temperature. The lateral transport of the thermally excited magnons, however, was recently investigated at both room and low temperatures using a nonlocal geometry \cite{cornelissen_long-distance_2015,giles_long-range_2015,cornelissen_temperature_2016}. In both studies relatively long magnon diffusion lengths have been found, one order of magnitude longer than reported in Ref. \cite{kehlberger_length_2015}. A YIG thickness-dependent study of the nonlocal thermal magnon transport is thus necessary to further clarify these issues.

Another way to generate incoherent magnons is spin-flip scattering with a non-equilibrium spin accumulation adjacent to the magnetic material \cite{xiao_transport_2015,bender_electronic_2012,takahashi_spin_2010}, for instance, in a spin Hall metal like Pt. A charge current through Pt creates a transverse spin current by the spin Hall effect (SHE), resulting in a spin accumulation at the YIG/Pt interface. Through interfacial exchange interaction, the angular momentum of the conduction electrons is transferred to the magnon system in YIG and thus creating or annihilating magnons, when the orientation of the spin accumulation is parallel or anti-parallel to the YIG order parameter. This electrical magnon injection method was first experimentally demonstrated to heat or cool the YIG lattice by magnon-phonon interaction, known as the spin Peltier effect \cite{flipse_observation_2014}. Recently, Cornelissen \textit{et al.} \cite{cornelissen_long-distance_2015} investigated the transport properties of such magnons using a lateral nonlocal geometry, with another Pt strip serving as a detector. This work demonstrates that incoherent magnons created electrically can also be used as an information carrier on a relatively long length scale, typically about 10 $\mu$m. Later this effect was compared with the spin Hall magnetoresistance (SMR) \cite{goennenwein_non-local_2015} and also observed in a vertical geometry \cite{li_observation_2016,wu_observation_2016}. In contrast to the auto-oscillation driven by the STT, this method was demonstrated to be a linear process \cite{flipse_observation_2014,cornelissen_long-distance_2015,li_observation_2016,wu_observation_2016} with respect to the injected current. Furthermore, this work is interpreted in terms of  non-equilibrium magnons, described by the magnon chemical potential \cite{cornelissen_magnon_2016}. For the results obtained on a 0.21 $\mu$m-thick YIG sample, the magnon propagation was well described in a diffusive model, driven by the magnon accumulation gradient. To further examine the magnon diffusive picture, the study for different YIG thicknesses is necessary.

In the device structure employed by Cornelissen \textit{et al.} \cite{cornelissen_long-distance_2015}, magnons are simultaneously excited both electrically and thermally, and the detection of these two types of magnons can be separated by the linear or quadratic dependence on the injection current. The magnons generated in these two methods exhibited very similar diffusion lengths, showing the same behavior in the long-distance regime. However, their short-distance behaviors are different, owing to the different magnon generation mechanisms. In this paper, by tuning the transparency of the YIG/heater interface from transparent to fully opaque for the spin currents, we associate the behavior of the magnons excited in these two ways also in the short distance regime, further proving their same nature. We also systematically investigate the effect of YIG thickness on the transport of electrically and thermally injected magnons, which allows us to examine the magnon diffusive transport model \cite{cornelissen_magnon_2016} and the bulk spin Seebeck model \cite{rezende_thermal_2014,duine_spintronics_2015}. 

This paper is organized as follows: Sec.~II presents the device configuration, fabrication details and measurement methods. In Sec.~III we first show the linear signals as a function of YIG thickness, where we probe the magnons that are injected electrically with a nonlocal geometry. Then we present the corresponding quadratic signals, which reflect the nonlocal behavior of the thermally generated magnons by the Joule heating in the injector. We show that the nonlocal SSE signals are strongly influenced by the transparency of the heater interface as well as the thickness of YIG. In Sec.~IV, we employ the two-dimensional finite element model (2D-FEM) and compare our experimental signals with the modeled results, and give an estimation of the bulk spin Seebeck coefficient. Finally, we discuss the deviations between the modeled and experimental results.

\section{II.~~~ Experimental details}

In our experiment, we used YIG (111) films with different thicknesses grown by LPE on single-crystal Gd$_{3}$Ga$_{5}$O$_{12}$(GGG) (111) substrates. The 0.21 $\mu$m, 1.5 $\mu$m, 12 $\mu$m and 50 $\mu$m-thick YIG samples were purchased from Matesy GmbH, and the 2.7 $\mu$m-thick YIG sample was provided by the Universit\'{e} de Bretagne in Brest, France. The FMR linewidths are similar among all the YIG samples ($<$ 2 Oe).

\begin{figure}[t]
	\includegraphics[width=8.5cm]{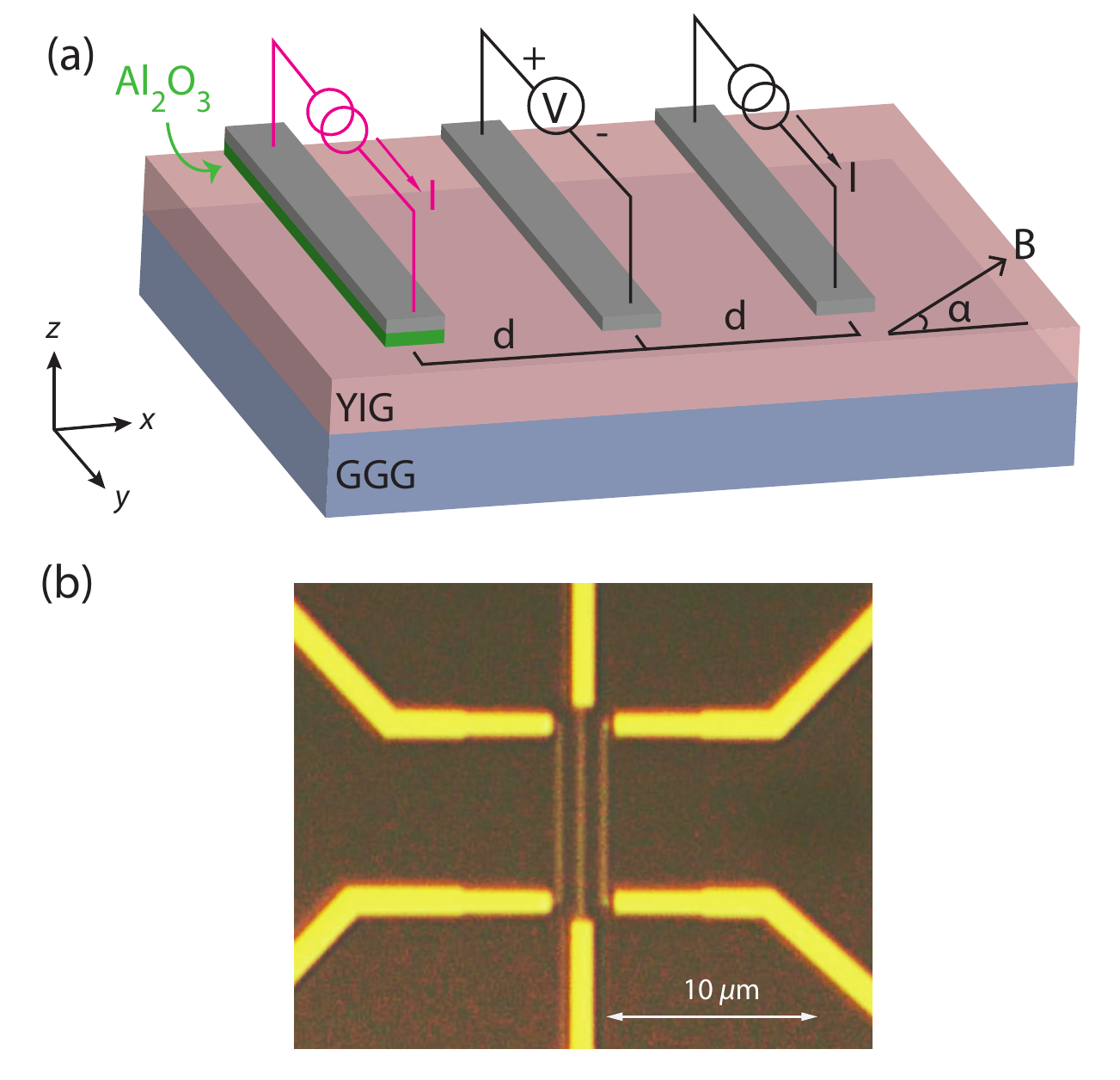}
	\caption{(a) Schematic representation of the device structure, where two Pt strips, with and without a thin (5 nm) Al$_2$O$_3$ layer underneath, are placed on the sides as injectors, and they share a Pt detector positioned in the middle. The center-to-center distance of the injector and detector is $d$, and $\alpha$ denotes the angle by which the in-plane magnetic field is applied. The Pt strips are all 7 nm in thickness. (b) The optical microscope image of one device, where the Pt strips are connected to Ti/Au contacts.}	
	\label{fig1}
\end{figure}

For each set of devices, three Pt strips that are 7 nm in thickness, typically with size 10 $\mu$m (length) $\times$ 100 nm (width), were sputtered at equal distance $d$ relative to each other. The device geometry is schematically shown in Fig.~\ref{fig1}(a). For the left strip, we deposited a thin Al$_2$O$_3$ layer (5 nm) by \textit{e}-beam evaporation before depositing Pt, in order to suppress the spin exchange interaction between Pt and YIG while preserving good thermal conduction. This provides a direct comparison to the right strip, where Pt is directly in contact with YIG. Equally large currents sent through both strips will generate the same Joule heating effects and the same temperature gradients in the YIG, and the only difference is the heater interface opacity for spin currents. Finally, the Pt strips were connected to Ti (5 nm)/Au (75 nm) contacts. We fabricated multiple sets of devices, with various heater-detector separation distances, ranging from 0.2 to 18 $\mu$m, on all our YIG samples. All structures were patterned using \textit{e}-beam lithography. For the long-distance device sets (where $d \geq$ 2 $\mu$m), we doubled the lengths of the Pt strips, in order to reduce the geometric effects so that the system can still be approximated to be a 2D problem in the $xz$ plane. The Pt widths were also increased accordingly, to allow for larger currents sent through and therefore boost the signal-to-noise ratio. The nonlocal results for these larger Pt strips were normalized carefully to the aforementioned typical size \footnote{For the electrically injected magnon detection, the $V^{1f}$ signals were first normalized by current, divided by the factor $(I/I_0)$, where $I$ is the used current and $I_0$ is the standard current 100 $\mu$A; and then normalized by Pt strip length, divided by the factor $(l/l_0)$, where $l$ is the used Pt strip and $l_0$ is standard length 10 $\mu$m. For the thermally excited magnon detection, the $V^{2f}$ signals were first normalized by current, divided by the factor $(I/I_0)^2$,  and then normalized to the Pt strip size, divided by the factor  $(l/l_0*w_0/w)$, where $w$ is the used Pt strip width and $w_0$ is the standard width 100 nm.}. 

For the measurements, we used a standard lock-in detection technique to separate the linear and quadratic effects, as described in our previous papers \cite{vlietstra_simultaneous_2014,shan_comparison_2015,cornelissen_long-distance_2015}. A low-frequency ($\sim$13 Hz) ac current, typically with an rms value $I_0$=100 $\mu$A, was sent through either the left or right strip, and the output voltage was nonlocally detected along the middle strip. The sample was rotated in a constant in-plane ($xy$ plane) magnetic field ($B=10$ mT), large enough to saturate the YIG magnetization \cite{vlietstra_spin-hall_2013}, and the signal was recorded as a function of the angle $\alpha$, as shown in Fig.~\ref{fig1}(a). The output voltage $V$ has both linear and quadratic contributions as $V=I_0 \cdot R_1 + I_0^2 \cdot R_2$, where $R_1$ and $R_2$ is the first and second order response coefficient, respectively, and is separated into the first ($V^{1f}$) and second ($V^{2f}$) harmonic signals by the lock-in measurement. When the third or even higher harmonic signals are negligible, as we checked is the case for our devices,  the first and second harmonic signals are proportional to $I_0$ and $I_0^2$, respectively \cite{vlietstra_simultaneous_2014,shan_comparison_2015,dejene_verification_2014}:
\begin{equation}
\begin{aligned}
& V^{1f}=I_0 \cdot R_1 \ \ \ \ \ \ \ \ \text{for} \ \ \ \phi =0^{\circ} \\
\text{and} \ \ 
& V^{2f}=\frac{1}{\sqrt{2}}I_0^2 \cdot R_2  \ \ \ \text{for} \ \ \  \phi =-90^{\circ}, 
\end{aligned}
\label{lock-in}
\end{equation}
where $\phi$ is the phase shift of the lock-in amplifier. $V^{1f}$ thus represents the linear signal where the non-equilibrium magnons are electrically injected via the SHE at the Pt injector, and detected nonlocally at the Pt detector via the ISHE; while $V^{2f}$ represents the quadratic spin Seebeck signal from Joule heating, where non-equilibrium magnons are thermally excited, and detected in the same fashion \cite{cornelissen_long-distance_2015}.

We also measured the locally generated voltage on the left (Pt/Al$_2$O$_3$) and right (Pt-only) strips. The local $V^{1f}$ is in this case the spin Hall magnetoresistance (SMR) signal \cite{nakayama_spin_2013,vlietstra_spin-hall_2013,chen_theory_2013} and $V^{2f}$ the local spin Seebeck signal induced by current heating \cite{schreier_current_2013,vlietstra_simultaneous_2014}. For the Pt/Al$_2$O$_3$ strips, the local $V^{1f}$ and $V^{2f}$ signals do not show any observable angular variations, indicating the effective suppression of the spin transport through the Al$_2$O$_3$ layer. For the Pt-only strips, the magnitudes of the SMR ratio ($\Delta R/R$) collected from different samples all fall in between $2 \times 10^{-4}$ and $3 \times 10^{-4}$. We can thus assume that the interface quality among our YIG samples is comparable. The local SSE results on Pt-only strips are shown in Appendix A. All measurements shown in this paper were performed at room temperature.

\section{III. ~~~Results and Discussion}

\subsection{A. ~~~Nonlocal results for electrically injected magnons}

\begin{figure}
	\includegraphics[width=8.5cm]{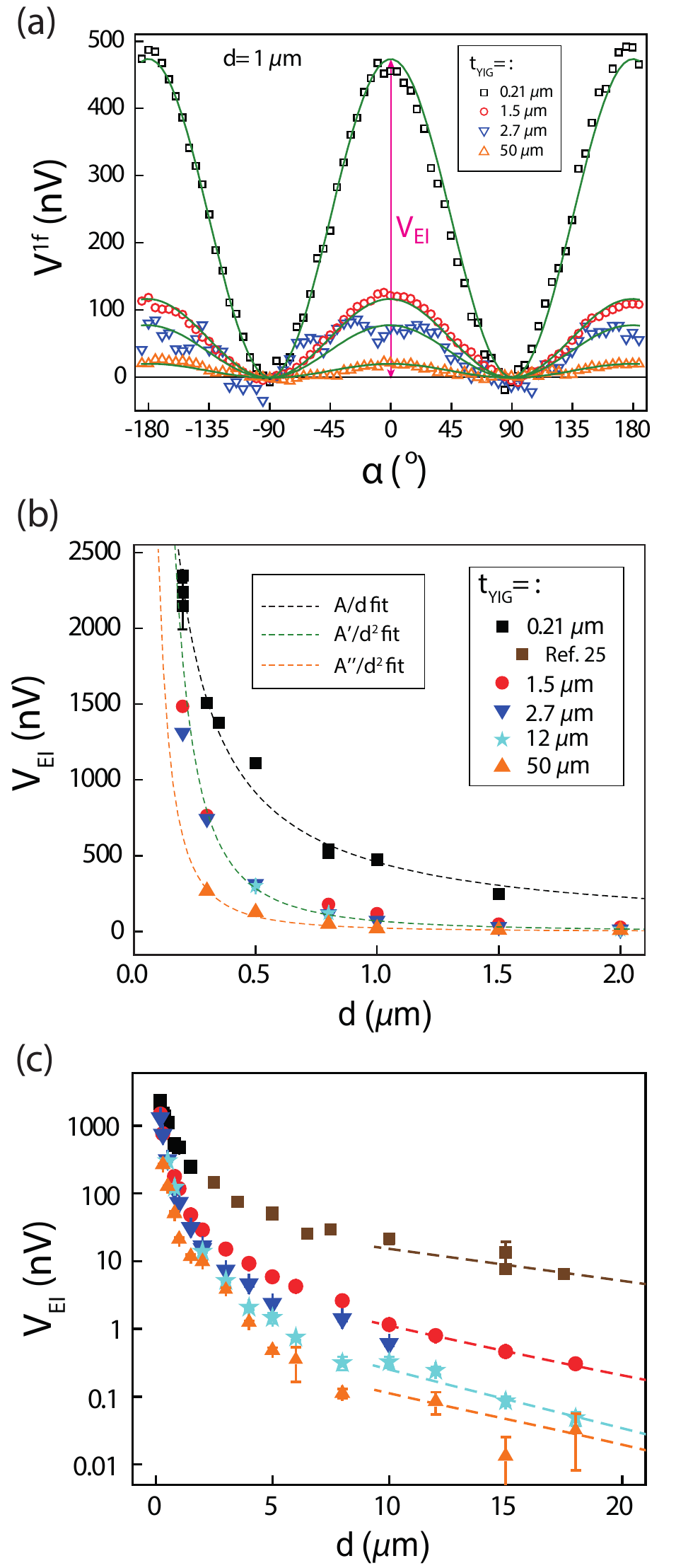}
	\caption{The first harmonic signal ($V^{1f}$) as a function of YIG thickness. (a) $V^{1f}$ as a function of $\alpha$, at the injector-detector spacing distance $d$=1 $\mu$m. The injected current $I$ has an rms value of 100 $\mu$A. The green solid curves are $\cos ^2 (\alpha)$ fits to the data. $V_{\textup{EI}}$ is defined as the amplitude of the electrically injected magnon signal. (b), (c), $V_{\textup{EI}}$ plotted as a function of $d$ for different YIG thicknesses, in linear ($d \leq$ 2 $\mu$m) and logarithmic scale, respectively. Dashed lines in (b) show the $A/d$ fit and $A'(A'')/d^2$ fits to the data. The data in brown squares in (c) are adapted from Ref.~\cite{cornelissen_long-distance_2015} for the sake of completeness. Dashed lines in (c) are the exponential fits using the parameters listed in Table~\ref{lambda}.}
	\label{fig2}
\end{figure}

We start by presenting the $V^{1f}$ results for various YIG thicknesses. Fig.~\ref{fig2}(a) shows the angular dependent results when using the right-side Pt-only strip as injector, with $d$=1 $\mu$m on different YIG samples. When sending a charge current through the injector, via the SHE a spin accumulation builds up at the bottom of the Pt strip, and its projection on the YIG magnetization will induce non-equilibrium magnons through the interfacial spin mixing conductance. The magnon injection efficiency is therefore proportional to cos($\alpha$), where $\alpha$ is the angle between the spin accumulation direction and the YIG magnetization. The injected magnons diffuse and at the same time relax in the YIG. When part of them successfully reach the detector, the reciprocal magnon detection process depends on cos($\alpha$) as well, and this in total gives a $\cos ^2 (\alpha)$ dependence. The signal thus reaches maximum $V_{\textup{EI}}$ when the spin accumulation in Pt is fully (anti)parallel with the external magnetic field ($\alpha$ = $-180^{\circ}, 0^{\circ} \  \textup{and} \ 180^{\circ}$), and we denote $V_{\textup{EI}}$ as the $V^{1f}$ signal amplitude.

It can be seen from Fig.~\ref{fig2}(a) that $V_{\textup{EI}}$ decreases as YIG becomes thicker, at the spacing distance $d$ = 1 $\mu$m. As we further plot $V_{\textup{EI}}$ as a function of $d$ for all YIG samples, as shown in Fig.~\ref{fig2}(b) and (c), we find that $V_{\textup{EI}}$ decreases monotonically as the YIG thickness increases, for nearly all spacings $d$. Particularly, for YIG thicker than 0.21 $\mu$m, $V_{\textup{EI}}$ decays faster as a function of $d$ in the short-distance regime. For a clear visualization we only plotted up to 2 $\mu$m in the linear scale in Fig.~\ref{fig2}(b). While for 0.21 $\mu$m $V_{\textup{EI}}$ exhibits a $1/d$ behavior, as we reported previously \cite{cornelissen_long-distance_2015}, for thicker YIG, $V_{\textup{EI}}$ no longer follows the $1/d$ behavior and can be better fitted with $1/d^2$ functions. 

As $d$ becomes larger, the $V_{\textup{EI}}$ signals can be better described by exponential decays, as can be seen in Fig.~\ref{fig2}(c). Similar slopes of $V_{\textup{EI}}$ as a function of $d$ can be observed, which indicates comparable $\lambda_m$ for all our YIG samples. We take the data points where $d >$ 8 $\mu$m for exponential decay fits and extract the $\lambda_m$ for different YIG samples, listed in Table.~\ref{lambda}. Given that $d =$ 8 $\mu$m may not yet be the onset for pure exponential decay, and that the $V_{\textup{EI}}$ signals for large $d$ gives larger uncertainties, the estimate of $\lambda_m$ from this method can be inaccurate. Nevertheless, the estimates in Table.~\ref{lambda} can be regarded as the lower limits of $\lambda_m$, as the pure exponential decays may start at a distance even further, which we could not probe due to reaching the noise limit of our detection method. We can conclude that the variance of $\lambda_m$ is not more than 50$\%$ among our samples; in fact, the variance could be actually smaller given the uncertainty from our estimation method. The reduction of the $V_{\textup{EI}}$ signals for thicker YIG samples, hence, can not be attributed to the different magnon spin relaxation lengths among our YIG samples. 

\begin{table}[t] 
	\caption{The estimated magnon diffusion length $\lambda_m$ for different YIG samples. Only the data points where $d >$ 8 $\mu$m were used for exponential fits, with the equation $V_{\textup{EI}}=A \cdot \exp (-d/\lambda)$, where $A$ is a coefficient that depends on YIG thickness. Given the large uncertainties in the datapoints on 50 $\mu$m YIG sample, the fitting weights were set to be larger for datapoints with smaller error bars.}
	\begin{ruledtabular}
		\begin{tabular} {c p{3cm}}
			YIG thickness ($\mu$m) & $\lambda_m$ ($\mu$m) \\
			\hline
			0.21 & 9.2 $\pm$ 1.0\\
			1.5 & 6.0 $\pm$ 0.3 \\
			2.7 & - \\
			12 & 5.0 $\pm$ 0.8  \\
			50 & 5.7 $\pm$ 3.4 \\
		\end{tabular}
	\end{ruledtabular}
	\label{lambda}
\end{table}

These observations cannot be fully explained by the magnon diffusive model \cite{cornelissen_magnon_2016}. From the diffusive picture, if the YIG thickness is increased, but is still much thinner than the magnon diffusion length $\lambda_m$, an increase of the $V_{\textup{EI}}$ would be expected, since from the injector to the detector the magnon channel is widened and hence the magnon conductance is increased. Magnon relaxation in the vertical $z$ direction enters when the YIG thickness becomes comparable to $\lambda_m$, in this case in the order of 9 $\mu$m. Increasing the YIG thickness even further would lead to a decrease of the signal, as the relaxation starts to play a more dominant role. This dependence has been calculated using the 2D-FEM with a magnon diffusion-relaxation model, as shown in Sec.~IV. In contrast, in our experiment $V_{\textup{EI}}$ reduces monotonically as the YIG thickness increases from 0.21 $\mu$m to 50 $\mu$m. Also, the stronger-decay behavior in the short-distance regime for thicker YIG samples cannot be fully explained.

When using the left-side Pt/Al$_2$O$_3$ strip as injector, the $V^{1f}$ signals do not show any observable angular dependences, as expected. This further confirms that the spin current through the YIG/Pt interface indeed plays a crucial role in this linear effect, and that the interface becomes fully opaque with a thin Al$_2$O$_3$ layer inserted in between.

 \begin{figure*}[t]
 	\includegraphics[width=16cm]{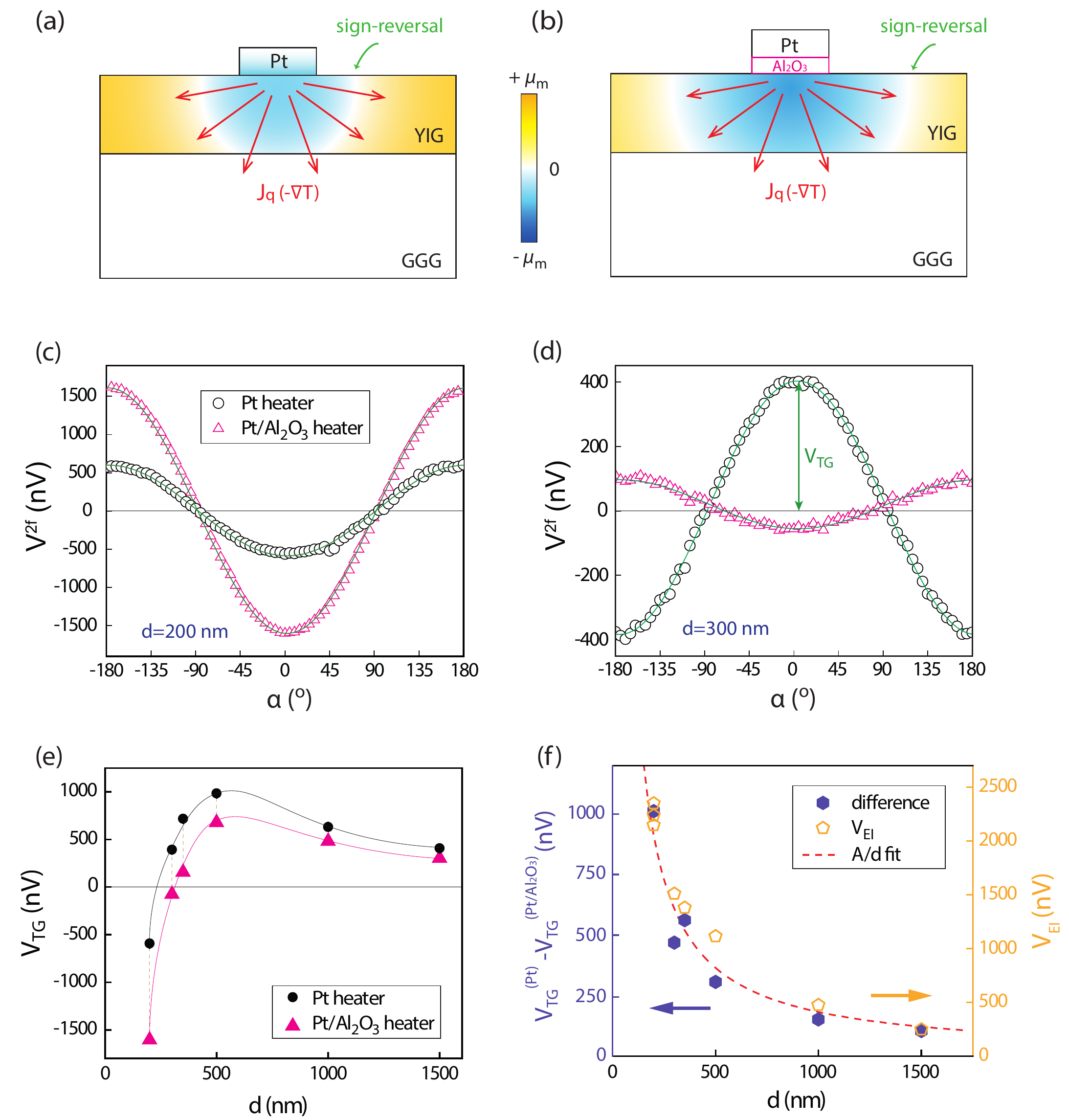}
 	\caption{The nonlocal detection of the thermally generated magnons for 0.21 $\mu$m YIG sample. (a), (b), Cross-section view of the magnon accumulation $\mu_m$ profile under a radial temperature gradient, when current is sent through the Pt heater (a) or Pt/Al$_2$O$_3$ heater (b), respectively. Red arrows represent heat flows $J_q$, and + (-) $\mu_m$ denotes magnon accumulation (magnon depletion), in color yellow (blue). (c), (d), Second harmonic signal $V^{2f}$ as a function of $\alpha$, with an rms injection current of 100 $\mu$A. In these plots the heater-detector distance is 200 nm and 300 nm, respectively. The black circles and pink triangles show results when the current is sent through either the Pt-only or the Pt/Al$_2$O$_3$ strip. Solid green curves are the cos($\alpha$) fits. $V_{\textup{TG}}$ are defined as the amplitude of the thermally excited, nonlocal SSE signal. (e) $V_{\textup{TG}}$ as a function of the heater-detector distance for both heating configurations. Solid curves are guidelines for the eyes. (f) The difference of the $V_{\textup{TG}}$ between the two heating configurations (solid purple polygons) compared with the electrically injected signal $V_{\textup{EI}}$ (open yellow polygons). Both of them follow the $1/d$ behavior.}
 	\label{fig3}
 \end{figure*}

\subsection{B. ~~~Nonlocal results for thermally generated magnons}

\subsubsection{1. ~~~The effect of the heater interface transparency}

Now we move to the $V^{2f}$ results, which represent the nonlocal signals of the thermally generated magnons, or the nonlocal SSE. The Joule heating effect of the injected current through the injector creates a radial temperature gradient in the YIG and GGG substrates, as shown in Fig.~\ref{fig3}(a)(b). Firstly, in this subsection, we show the strong influence of the heater interface transparency on nonlocal SSE signals by comparing the results between sending currents through the Pt-only strip and the Pt/Al$_2$O$_3$ strip. The temperature profiles of these two heating configurations are very comparable, given that the Pt strips are identical and that the Al$_2$O$_3$ layer is thin (5 nm). It has been checked in the 2D-FEM that the temperature profile ($T-T_0$, where $T$ is the lattice temperature and $T_0$ is the room temperature) varies not more than 3$\%$ locally and 0.02$\%$ nonlocally with the insertion of the Al$_2$O$_3$ layer (see Appendix A). 

The results for the device sets on the 0.21 $\mu$m YIG sample are presented in Fig.~\ref{fig3}. Fig.~\ref{fig3}(c) shows the angular dependence for the measured $V^{2f}$ when $d$ is 200 nm, for both the two heating configurations where the current is sent through the Pt/Al$_2$O$_3$ or Pt strip. Both curves show a $\textup{cos} (\alpha)$ behavior, which is governed by the ISHE at the detector. Strikingly, for the same distance, same heating power, the $V^{2f}$ signals for the two heating configurations differ by a factor of three. Even more interestingly, when $d$ is 300 nm, the $V^{2f}$ signals of the two heating configurations show opposite signs, as shown in  Fig.~\ref{fig3}(d). Given that the only difference between the two configurations is the heater transparency, it can be inferred that the thermally generated magnon flow does not only rely on the temperature profile, but is also sensitive to the heater opacity at some distance away. 

The difference between the two heating configurations can be seen more clearly in the distance dependence data. We define $V_{\textup{TG}}$ as the magnitude of $V^{2f}$, and plot it for both heating configurations as a function of $d$ in Fig.~\ref{fig3}(e). Note that the negative sign of $V_{\textup{TG}}$ corresponds to the same sign as the SSE signal measured locally. For the Pt heater series, a sign reversal of the $V_{\textup{TG}}$ occurs when $d$ is in between 200 and 300 nm, consistent with the results we reported in Ref.~\cite{cornelissen_long-distance_2015}, though in this study the YIG sample is from a different provider. For the other Pt/Al$_2$O$_3$ heater series, the sign reversal of $V_{\textup{TG}}$ occurs at a slightly further distance, between 300 and 350 nm. In fact, for each $d$, the signals obtained from heating the Pt/Al$_2$O$_3$ strip are always more negative than for heating the Pt-only strip. These results strongly indicate that the thermally generated magnon current is not only determined by the temperature profile, but also sensitive to the boundary conditions that modify the magnon currents.

These observations can be described by the concept of a bulk SSE theory \cite{rezende_magnon_2014,duine_spintronics_2015,cornelissen_magnon_2016}. An analytical description can be found in Appendix C. According to this theory, a heat flow $J_q$ in YIG will excite a thermal magnon flow $J_{m,q}$ along with it, related by the bulk spin Seebeck coefficient $S_S$:
\begin{equation}
J_{m,q}=- \sigma_m S_S \nabla T~\propto J_q=-\kappa \nabla T,
\label{eq:Jmq}
\end{equation}
where $\sigma_m$ is the magnon conductivity and $\kappa$ is the thermal conductivity of YIG. While the heat flow is continuous through the boundaries, the magnon flow stops, resulting in the built-up of magnon accumulations $\mu_m$, opposite in sign for the YIG/heater and YIG/GGG boundaries. The positive $\mu_m$ corresponds to more magnons in excess of equilibrium, hence magnon \textit{accumulation}; and the negative $\mu_m$ corresponds to fewer magnons as compared to equilibrium, hence magnon \textit{depletion}. This picture is analogous to the traditional Seebeck effect in conductive systems, where a positive and negative charge voltages are built up as a result of a temperature gradient. 

A diffusive magnon flow $J_{m,diff}$ is induced to balance the thermal magnon flow, until the system reaches a steady state:
\begin{equation}
J_{m,diff}=- \sigma_m \nabla \mu_m.
\label{eq:Jdiff}
\end{equation}
The total magnon current ($J_m=J_{m,diff}+J_{m,q}$) hence includes both the thermal and diffusive parts, and relaxes on the length scale of $\lambda_m$:  
\begin{equation}
\boldsymbol{\nabla}\cdot\boldsymbol{\mathit{J_m}}=-\sigma_m\frac{\mu_m}{\lambda_m^2}.
\label{eq:Jtotal}
\end{equation}

In our device geometry, owing to the radial temperature gradient, an intensive negative $\mu_m$ builds up beneath the heater, surrounded by the sparsely distributed positive $\mu_m$ (supposing a positive $S_S$), as shown in Fig.~\ref{fig3} (a)(b). When placing a Pt detector nonlocally at the YIG surface, the Pt detector then serves as a spin sink, extracting or injecting a certain magnon flow, depending on the sign of the $\mu_m$ at that position. The nonlocal signal would hence first probe the negative $\mu_m$ for shorter $d$ and then the positive $\mu_m$ for longer $d$, reversing sign in between. 

Changing the transparency of the YIG/heater interface will influence the amount of negative $\mu_m$ below the heater, and thus tune the sign-reversal distance. Compared to the fully opaque YIG/heater interface for the Pt/Al$_2$O$_3$ heater series, the transparent YIG/Pt interface allows for certain magnon flow into the heater via the spin mixing conductance, hence a less negative $\mu_m$ will be preserved beneath the heater. Consequently, the sign-reversal occurs at a shorter $d$, closer to the heater (see Fig.~\ref{fig3}(a)). The fully opaque interface thus corresponds to the furthest sign-reversal distance, as shown in Fig.~\ref{fig3}(b). Our results confirm the fact that, in additional to the temperature profile, the magnon accumulation and the magnon current are essential in the spin Seebeck picture.

\begin{figure*}[t]
	\includegraphics[width=16cm]{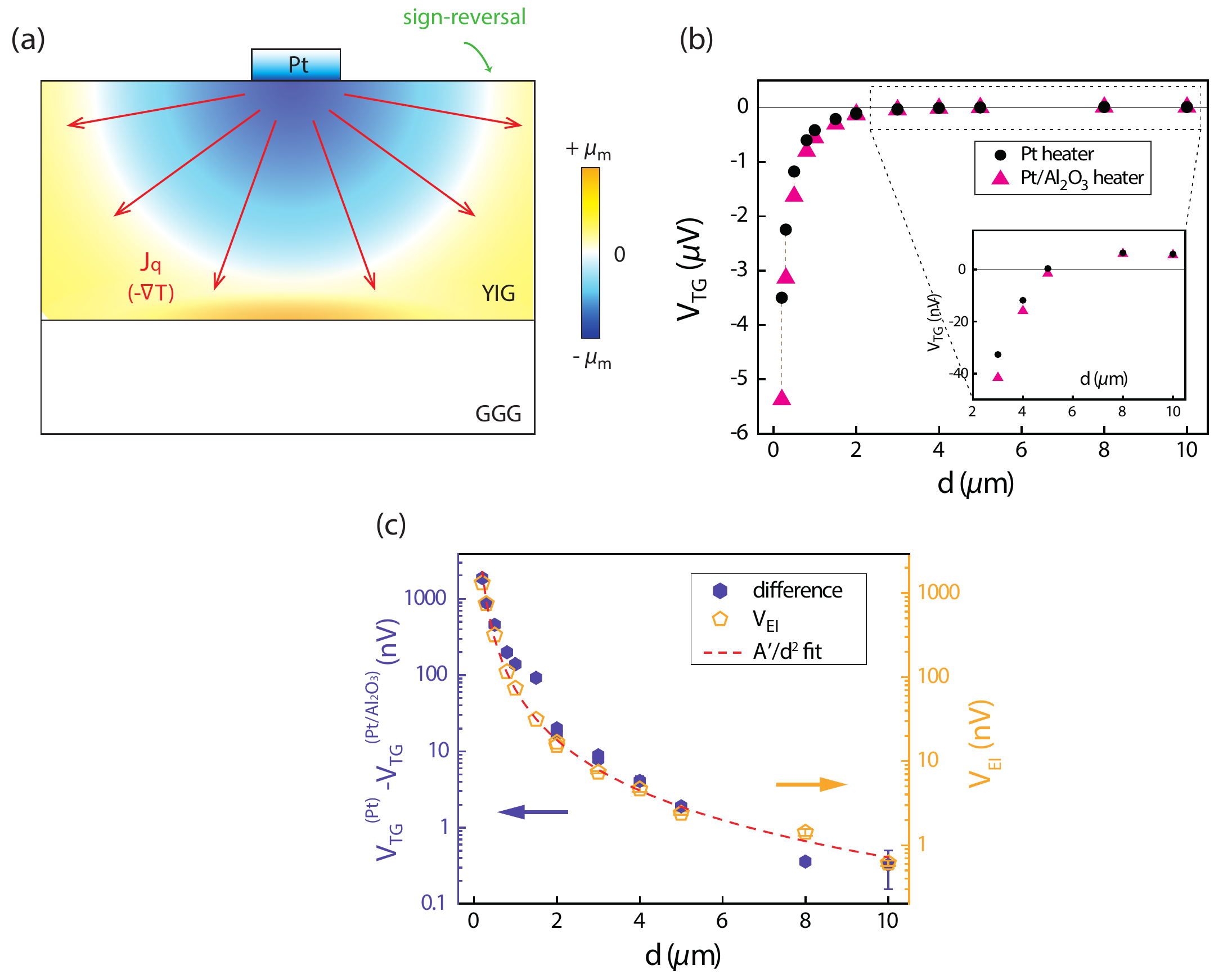}
	\caption{The nonlocal detection of the thermally generated magnons for 2.7 $\mu$m YIG sample. (a) Cross-section view of the magnon accumulation $\mu_m$ profile under a radial temperature gradient, when the YIG thickness is increased. Larger magnon accumulations are present at both the YIG/heater and YIG/GGG boundaries, compared to the situation for thinner YIG. (b) $V_{\textup{TG}}$ as a function of the heater-detector distance for both heating configurations. Inset zooms in for longer-distance data and shows the sign-reversal behavior. (c) The difference of $V_{\textup{TG}}$ between the two heating configurations compared with the electrically injected signal $V_{\textup{EI}}$, plotted in logarithmic scale.}
	\label{fig4}
\end{figure*}

Remarkably, the difference of the signals from the two heating configurations exhibits a $1/d$ behavior, similar to the electrical injection induced signal ($V_{\textup{EI}}$) shown in the previous section, as plotted in Fig.~\ref{fig3}(f). This can also be explained by the bulk SSE picture: in comparison with the Pt heating series, the Pt/Al$_2$O$_3$ heating series has an extra negative $\mu_m$ beneath the heater. It can be compared with the non-equilibrium magnons created by electrical injection at the injector. The fact that both of them can be fitted to a $1/d$ behavior suggests that magnons generated thermally and electrically are very similar in nature. 

At first sight, our results could be reminiscent of the transverse SSE experiments performed by Uchida \textit{et al.} \cite{uchida_spin_2010} with the sign-reversal feature. It is important to point out a fundamental difference between the two experiments: in our experiment the spatial variation of $\mu_m$ can only be observed a few times of $\lambda_m$  away from the heater, whereas in Ref.~\cite{uchida_spin_2010} the SSE signal is varying throughout the whole YIG in the range of a few millimeters, which cannot be explained in the magnon diffusive framework with the so-far reported $\lambda_m$ in YIG \cite{boona_magnon_2014,kehlberger_length_2015,cornelissen_long-distance_2015,giles_long-range_2015}. Our results hence do not share the same origin as the transverse SSE.

\subsubsection{2. ~~~The effect of the YIG thickness}

Apart from the transparency of the YIG/heater interface, varying the YIG thickness is also expected to influence the nonlocal spin Seebeck signals, due to the bulk nature of the SSE \cite{kehlberger_length_2015,rezende_magnon_2014,duine_spintronics_2015}. Fig.~\ref{fig4} shows the measured $V_{\textup{TG}}$ results on a 2.7 $\mu$m-thick YIG sample. As can be immediately seen, the distance-dependences of $V_{\textup{TG}}$ (Fig.~\ref{fig4}(b)) of both heating configurations have very different shapes as compared with the 0.21 $\mu$m YIG sample (Fig.~\ref{fig3}(d)). Again the negative sign corresponds to the sign of the local SSE signal. For the 2.7 $\mu$m YIG, the sign-reversals for both heating series take place much further, around 5 $\mu$m as shown in the inset. In addition, for the very short distances, as when $d$= 200 nm, the SSE signals of the thicker YIG are a few times larger compared with the thinner YIG, for both heating configurations. It is interesting to point out that the local SSE signals we measured on the Pt-only strips do not show such a big difference between the 0.21 $\mu$m YIG and 2.7 $\mu$m YIG (see Appendix B for more discussion). 

The different behavior of $V_{\textup{TG}}$ with varying YIG thickness can be understood as following: when YIG becomes thicker, the positive and negative $\mu_m$ will be separated further and have a smaller counter effect to each other. As a result, both the positive and negative $\mu_m$ will increase, and the positive $\mu_m$ will be pushed further away from the heater, more sparsely distributed at a larger YIG volume, as shown in Fig.~\ref{fig4}(a). Therefore, the sign-reversal distance becomes larger as the YIG thickness increases.

One common feature is observed for both 0.21 and 2.7 $\mu$m YIG samples: for all distances, the signals from the Pt/Al$_2$O$_3$ heater series are more negative than the Pt-only heater series. For the 2.7 $\mu$m YIG sample, we can also plot the difference between the two heater series as a function of $d$, shown in Fig.~\ref{fig4}(c). Its shape matches with the $V_{\textup{EI}}$ signal, both can also be described by a $1/d^2$ behavior. This observation proves again the similar nature for the electrically and thermally excited magnons.

More results from other YIG samples with different thicknesses are shown in Fig.~\ref{fig5}, in logarithmic scale (Plots in linear scale can be found in Appendix D). In this plot we include the results for a third measurement configuration: sending current through the Pt/Al$_2$O$_3$ heater, and measuring voltage at the right Pt strip, which in this case serves as the detector. This measurement configuration enables us to probe twice as far distance data for our present devices, and investigate the effect of a Pt absorber (the middle Pt strip) in between the heater and detector for nonlocal SSE. Comparing the results from this configuration (star-shaped symbols) and the Pt/Al$_2$O$_3$ heater series in Fig.~\ref{fig5}, we can conclude that there is only a small reduction, mostly within 10$ \%$, when there is a Pt absorber present in between. It is therefore reliable to include this series to look at how the $V_{\textup{TG}}$ decay as a function of $d$ for the long-distance regime. It can be seen from Fig.~\ref{fig5} that for all YIG samples the exponential decay rates are comparable. Using the data points where $d >$ 8 $\mu$m in the exponential fits, we obtain $\lambda_m$ of 7.5 $\pm$ 0.5 $\mu$m for the 1.5 $\mu$m YIG sample and 6.1 $\pm$ 0.4 $\mu$m for the 50 $\mu$m YIG sample. Comparing with the 0.21 $\mu$m sample which gives a $\lambda_m$ of 9.6 $\pm$ 1.0 $\mu$m, this further proves the fact that for long-$d$ regime, $\lambda_m$ is not varying by more than 30\% among different thick YIG samples. 

We can also plot the sign-reversal distance as a function of YIG thickness, as shown in the inset of Fig.~\ref{fig5}. As expected, the sign reversal takes place at a further distance for thicker YIG. For the 50 $\mu$m YIG sample we could not observe the sign reversal for the distance range we investigated. The trend can be fitted to a linear dependence, and the sign-reversal distance is around 1.6 times the YIG thickness.

\begin{figure}[t]
	\includegraphics[width=9cm]{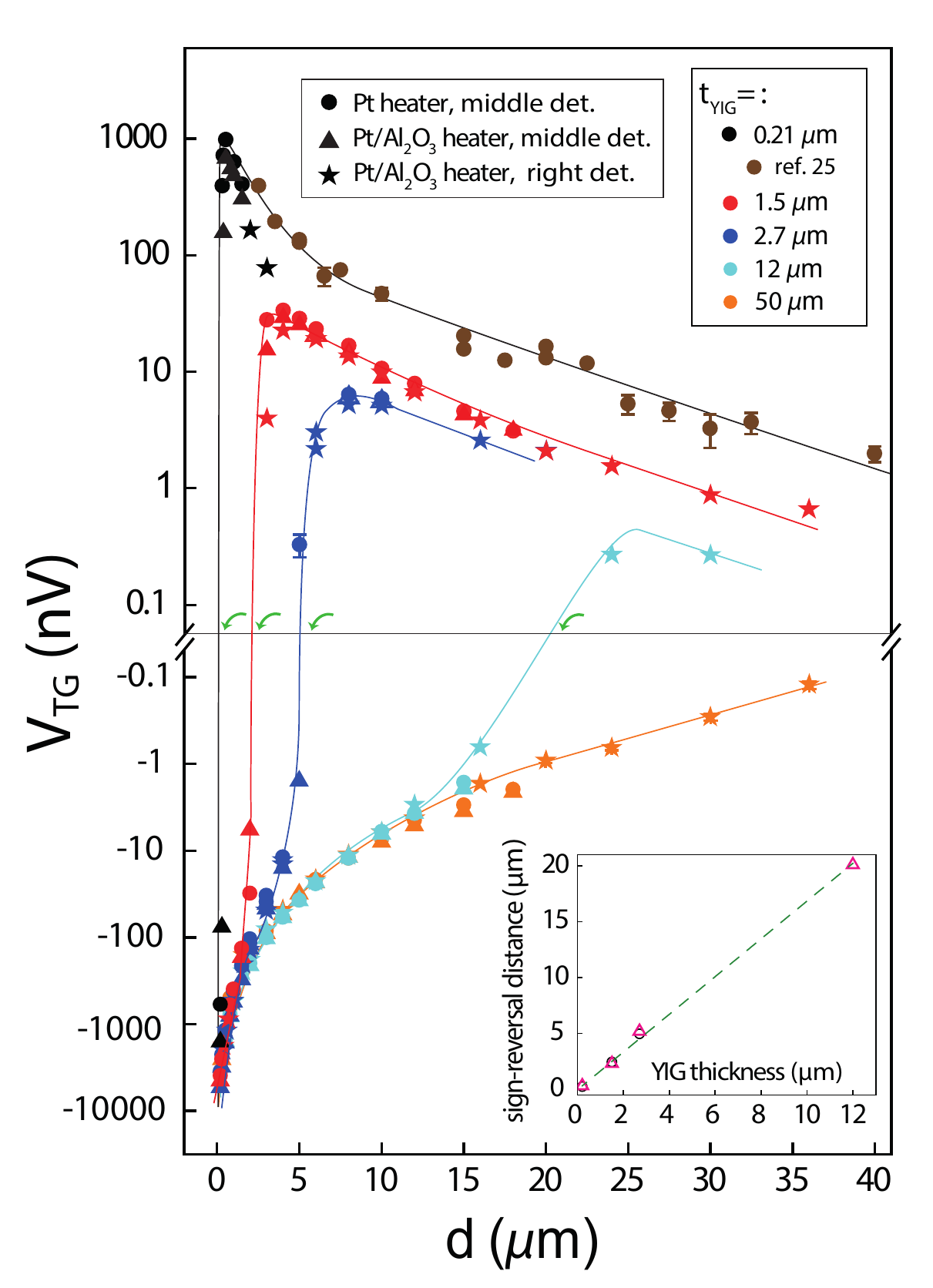}
	\caption{Nonlocal $V_{\textup{TG}}$ results as a function of $d$ for different YIG thicknesses (indicated by symbol colors) and heating configurations (indicated by symbol shapes), plotted in a logarithmic scale. The data from the third heating configuration, where current is sent through the left Pt/Al$_2$O$_3$ strip and voltage is measured at the right Pt strip, are shown in this figure with star shaped symbols for all YIG samples. The brown circles are adapted from Ref.~\cite{cornelissen_long-distance_2015} for the sake of completeness. Solid curves are guidelines for the eyes, and green arrows indicate sign reversals. Inset plots the sign-reversal distance as a function of the YIG thickness.}
	\label{fig5}
\end{figure}

\section{IV. ~~~Finite Element Modeling Results}

Using a 2D steady-state FEM allows us to quantitatively compare our results with the theory. In this section, we present the 2D-FEM results for the nonlocal behavior of the electrically and thermally injected magnons in the framework of a pure magnon diffusive model \cite{cornelissen_magnon_2016}, where the magnon current is driven by the non-equilibrium magnon accumulation $\mu_m$.

\subsection{A. ~~~Electrically injected magnons}

\begin{table}[t]
	\caption{Material parameters that were used in the model. $\sigma_{e}$ and $\sigma_{s}$ ($\sigma_{m}$) is the electron and spin (magnon) conductivity, respectively. For the YIG/Pt interface, the spin conductivity $\sigma_m$ is calculated by $\sigma_m = g_S \cdot t_{\textup{interface}}$, where $g_S$ is the effective spin mixing conductance \cite{flipse_observation_2014}, and was estimated in our recent work \cite{cornelissen_magnon_2016}. The other parameters of the YIG/Pt interface are assigned to be the same as YIG. Note that the spin conductivity of a paramagnetic metal, such as Pt, is half of its electrical conductivity \cite{chen_theory_2013}. The spin Hall angle of Pt $\theta_\textup{SH}$ is taken as 0.11 \cite{schreier_magnon_2013,flipse_observation_2014,cornelissen_magnon_2016}.}
	\begin{ruledtabular}
		\begin{tabular}{c c c c c c}
			\renewcommand{\arraystretch}{2}
			Material & $\sigma_{e}$ & $\sigma_{s}$ ($\sigma_{m}$)  & $\kappa$ & $\lambda$ \\
			(thickness) & (S/m) & (S/m)  & (W/(m$\cdot$K)) & (m) \\
			\hline
			Pt (7 nm) & 2.5$\cdot$10$^6$  & 1.25$\cdot$10$^6$ & 26 & 1.5 $\cdot$10$^{-9}$\\
			YIG/Pt interface (1 nm) & - & 0.96$\cdot$10$^4$ & 6 & 9.4$\cdot$10$^{-6}$\\
			Al$_2$O$_3$ (5 nm) & - & - & 0.15 & - \\
			YIG (various thickness) & - & 5$\cdot$10$^5$  & 6 & 9.4$\cdot$10$^{-6}$ \\
			GGG (500 $\mu$m) & - & - & 8 & - \\
		\end{tabular}
	\end{ruledtabular}
	\label{parameter}
\end{table}

\begin{figure}[t]
	\includegraphics[width=8cm]{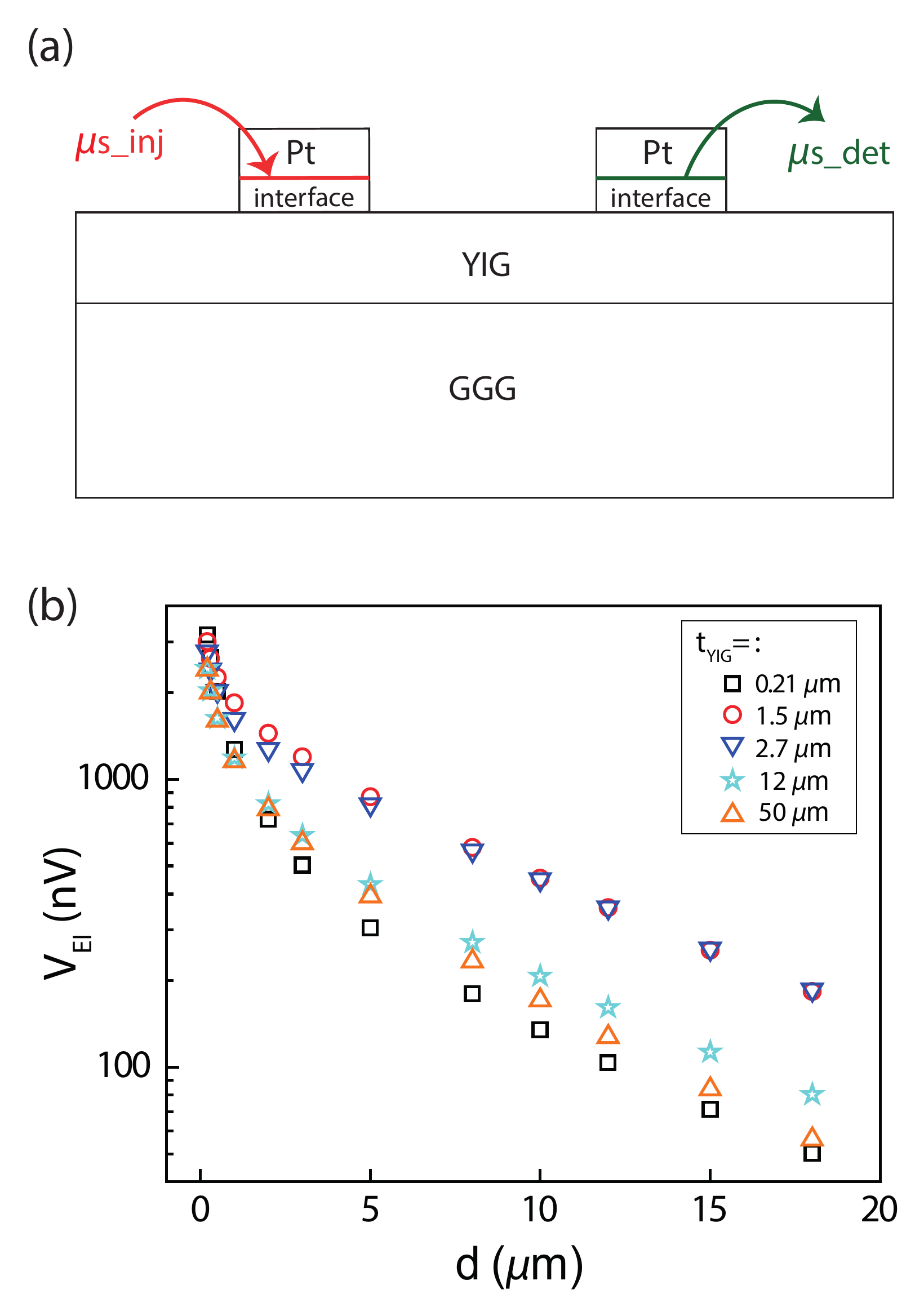}
	\caption{The calculated $V_{\textup{EI}}$ results as a function of $d$ for different YIG thicknesses. (a) Schematic illustration of geometry that was employed in the model. The injected spin voltage $\mu_{\textup{s}\_\textup{inj}}$ is set as a Dirichlet boundary condition and the spin voltage at the detector $\mu_{\textup{s}\_\textup{det}}$ is extracted from the calculation. (b) The modeled $V_{\textup{EI}}$ results plotted on a logarithmic scale.}
	\label{fig6}
\end{figure}

First we discuss the transport of the electrically injected magnons. The model solves in the whole geometry the magnon (spin) transport equation
\begin{equation}
\boldsymbol{\mathit{J_m}} =-\sigma_m\cdot\boldsymbol{\nabla} \mu_m,
\label{eq:jm}
\end{equation}
where $\boldsymbol{\mathit{J_m}}$ is the magnon current density, $\sigma_m$ is the magnon spin conductivity and $\mu_m$ is the magnon (spin) accumulation. The relaxation of the magnons is described by the Valet-Fert equation \cite{valet_theory_1993,zhang_magnon_2012}
\begin{equation}
{\nabla}\cdot \boldsymbol{\mathit{J_m}}=-\sigma_m\frac{\mu_m}{\lambda_m^2}.
\end{equation}
This equation is applied to the whole geometry shown in Fig.~\ref{fig6}. The interface is modeled as a layer with thickness $t_{\textup{interface}}$ equal to 1 nm \cite{cornelissen_magnon_2016}. The spin conductivity of the interface is then $g_S \cdot t_{\textup{interface}}$, where $g_S$ is the effective spin mixing conductivity \cite{flipse_observation_2014,xiao_transport_2015,cornelissen_magnon_2016}.

The SHE and ISHE processes in the Pt are not included in the model, but calculated analytically. The spin accumulation at the bottom of Pt created by the SHE is denoted by $\mu_{\textup{s}\_\textup{inj}}$, and is calculated as \cite{flipse_observation_2014,chen_theory_2013} 
\begin{equation}
\mu_{\textup{s}\_\textup{inj}}=\frac{2 e}{\sigma_\textup{pt}}\cdot \lambda_\textup{pt}\cdot \theta_\textup{SH}\cdot J_c \cdot \tanh \left ( {\frac{t_\textup{pt}}{2\lambda_\textup{pt}}} \right),
\label{she}
\end{equation}
where $e$ is the electron charge, $t_\textup{pt}$, $\lambda_\textup{pt}$ and $\sigma_\textup{pt}$ is the thickness, spin diffusion length and electrical conductivity of Pt, respectively; $\theta_\textup{SH}$ is the spin Hall angle of the Pt, and $J_c$ is the injected electric charge current density, equal to $1.43 \times 10^{11}$ A/m$^2$. $\mu_{\textup{s}\_\textup{inj}}$ serves as the input of the model.

The output of the model is extracted from the spin accumulation $\mu_{\textup{s}\_\textup{det}}$ at the detector. Following the derivation from Ref. \cite{castel_platinum_2012}, The induced ISHE electrical voltage, which equals $V_{\textup{EI}}$ here, is expressed as
\begin{equation}
V_{\textup{ISHE}}=\frac{1}{2 e}\cdot \frac{L_\textup{pt}}{t_\textup{pt}}\cdot \theta_\textup{SH}\cdot \frac{(1-e^{-\frac{t_\textup{pt}}{\lambda_\textup{pt}}})^2}{1+e^{-\frac{2t_\textup{pt}}{\lambda_\textup{pt}}}}\cdot \mu_{\textup{s{\_}det}},
\label{eq:ishe}
\end{equation}
where $L_\textup{pt}$ is the length of the Pt strip. To be consistent with our previous calculations, for all parameters, we take the same values as used in Ref. \cite{cornelissen_magnon_2016}, except for the $\sigma_\textup{pt}$ which is $2.5 \times 10^6$ S/m extracted from the average Pt resistance from the measured Pt strips. The used material parameters are listed in Table.~\ref{parameter}.

The calculated results for different YIG thicknesses are shown in Fig.~\ref{fig6}(b). The modeled results do not show the same trend as the experimental results: except for the datapoints at very short $d$, the modeled signals in general increase first with increasing the YIG thickness, when the YIG thickness is still much smaller compared to $\lambda_m$. Further increase of the YIG thickness then decreases $V_{\textup{EI}}$, as the magnon relaxation in the vertical direction starts to play a role. This trend is different from the monotonic decrease of the $V_{\textup{EI}}$ with increasing the YIG thickness observed in experiment. Moreover, in the short-$d$ regime, the modeling results cannot capture the sharp decrease of the signals as observed experimentally for thicker YIG samples. 

These discrepancies between the modeling and experiments indicate the limits of a model based on magnon spin accumulation only, and may call for additional shorter length scales in the short-distance regime, such as the magnon-phonon and other relaxation lengths introduced in Ref.~\cite{cornelissen_magnon_2016}. Close to the injector the magnon diffusion may be characterized by a shorter length scale. This scenario can explain the significant drop of the $V_{\textup{EI}}$ from 0.2 $\mu$m to 1.5 $\mu$m YIG samples, as 0.2 $\mu$m is still within or comparable to this shorter length scale but 1.5 $\mu$m far excesses it, resulting in more magnon relaxation. The vertical relaxation thus begins at much thinner YIG than modeled. The faster decay of the $V_{\textup{EI}}$ in thicker YIG samples could also be understood when taking into account another shorter length scale. More discussions can be found in Sec.~IV-C.

\subsection{B. ~~~Thermally generated magnons}

\begin{figure*}[t]
	\includegraphics[width=14cm]{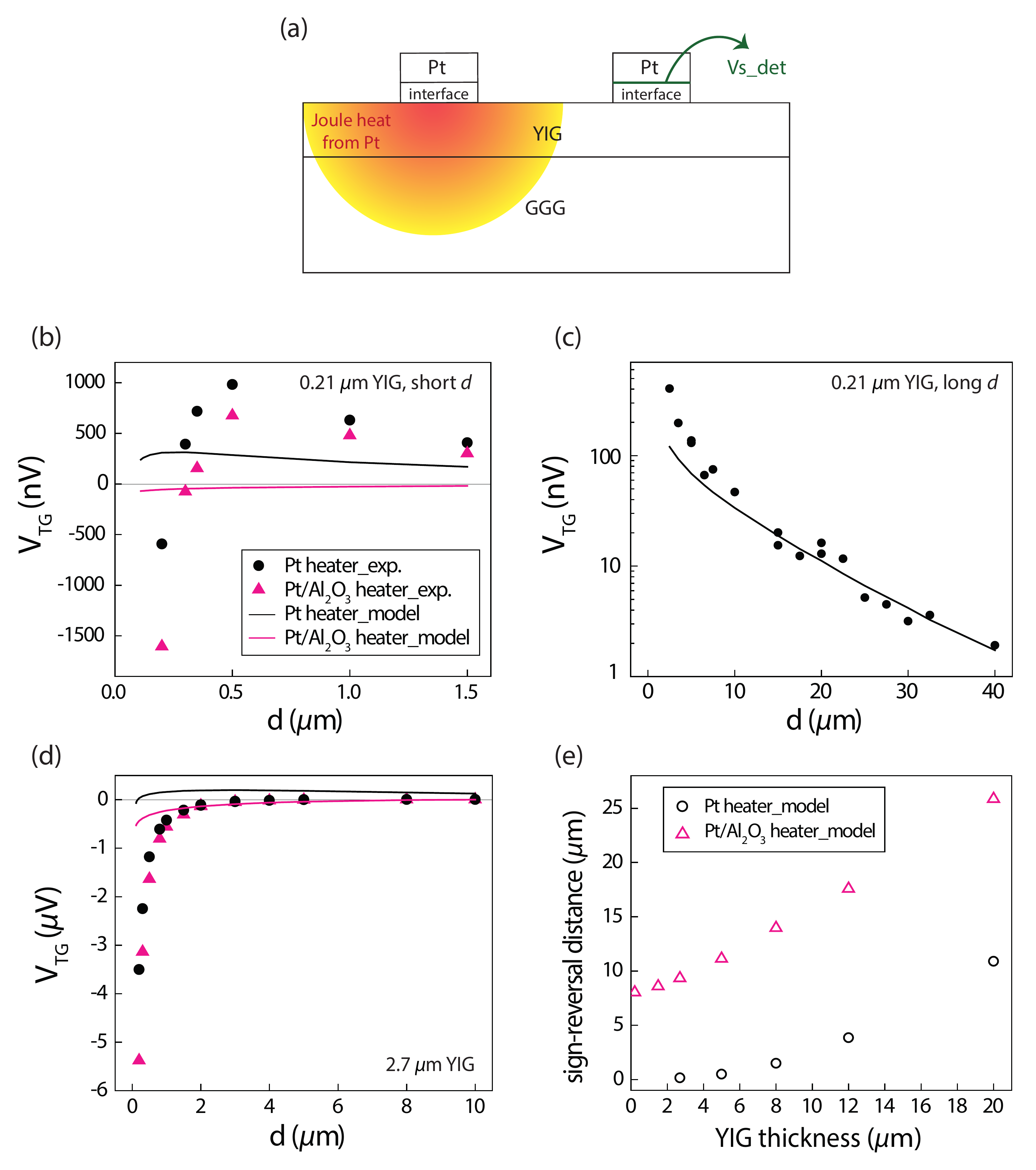}
	\caption{The modeling of nonlocal SSE signals with parameters in Table.~\ref{parameter} and $S_S$=4.5 $\mu$V/K. (a) The 2D geometry that was studied in the model. (b)(c), Modeled results (solid lines) compared with experimental (solid symbols) nonlocal  SSE signals for the two heating series on 0.21 $\mu$m YIG for short (b) and long (c) distances. The experimental data (black dots) in (c) are from Ref.~\cite{cornelissen_long-distance_2015}. (d), The comparison between experimental and modeled results for 2.7 $\mu$m YIG. (e), The calculated sign-reversal distances as a function of the YIG thickness for the two heating configurations.}
	\label{fig7}
\end{figure*}

We can also use the 2D-FEM to obtain a quantitative picture of the nonlocal behavior for the thermally generated magnons. 

We consider the magnon spin current flow and the heat flow, related to their driving forces as \cite{cornelissen_magnon_2016}:
\begin{equation}
\begin{pmatrix} \boldsymbol{\mathit{J_m}} \\ \boldsymbol{\mathit{Q}} \end{pmatrix} =- \begin{pmatrix} \sigma_m & \sigma_m S_S \\ \sigma_m S_ST & \kappa \end{pmatrix} \begin{pmatrix} \boldsymbol{\nabla} \mu_m \\ \boldsymbol{\nabla} T \end{pmatrix}
\label{eq:jq}
\end{equation}
where $S_S$ is the bulk magnon Seebeck coefficient that is only non-zero for YIG, and we assume it to be the same for different YIG thicknesses, as an intrinsic material parameter. The source terms of the two current flows are 
\begin{equation}
\boldsymbol{\nabla}\cdot\boldsymbol{\mathit{J_m}}=-\sigma_m\frac{\mu_m}{\lambda_m^2}\ \  \textup{and} \ \ 
\boldsymbol{\nabla}\cdot\boldsymbol{\mathit{Q}}=\frac{\mathit{J_c}^2}{\sigma_\textup{pt}},
\label{eq:jqsource}
\end{equation}
 where the first equation stands for the magnon relaxation, and the second equation represents the Joule heating effect. The Joule heating only takes place in the heater, and serves as the input in the spin Seebeck scenario. The output of the signal is also extracted from the $\mu_{\textup{s}\_\textup{det}}$ at the detector, from which the ISHE voltage is calculated using Eq.~\ref{eq:ishe}. 

The modeled results are shown in Fig.~\ref{fig7}, with $S_S$ taken as 4.5 $\mu$V/K for all YIG samples. The fitting for the long-$d$ range is satisfactory, where only the magnon diffusion and relaxation take place, and the $V_{\textup{TG}}$ exhibits pure exponential decay. From the Pt heater series on 0.21 $\mu$m YIG (Fig.~\ref{fig7}(b)), we can determine the value of $S_S$ to be 4.5 $\mu$V/K.

The short-$d$ data, however, only shows qualitative agreement with the experimental data. The signals from the Pt/Al$_2$O$_3$ heater series are more negative than from Pt heater series, and the sign-reversal distance takes place at a further $d$ than Pt heater series, consistent with the observation from the experiments. As the YIG thickness increases, the sign-reversal distances also shift to further distance. But in the model, for the parameters we used from Table.~\ref{parameter}, the difference for the two heating configurations is larger than in experiment. Compared to the experiment, the sign-reversal for the Pt series is much closer to the heater, and for the Pt/Al$_2$O$_3$ heater series is much further away. Also, from Fig.~\ref{fig7}(d) one can see that the fast decay of the $V_{\textup{TG}}$ signals in the short-$d$ regime cannot be captured by the model; same as the electrical injection, a short length scale may be needed to be introduced in the short-$d$ regime.

\subsection{C. ~~~Summary}

So far the model works in showing that there are indeed sign reversals when probing the thermally generated magnon signals nonlocally, and that this sign reversal is indeed influenced by both the YIG thickness and the heater opacity. Moreover, the signals from the Pt/Al$_2$O$_3$ heater series are more negative than from Pt heater series, which is qualitatively consistent with the experimental results. However, full quantitative agreement cannot be reached.

Here we provide some tentative explanations of the quantitative deviation between the model and experiments. First of all, in our model we only consider $\mu_m$ to describe the non-equilibrium magnons, and assumes the magnon temperature $T_m$ to be the same as the phonon temperature $T_{ph}$, based on the very short magnon-phonon relaxation length \cite{flipse_observation_2014,cornelissen_magnon_2016,agrawal_direct_2013}. It could be possible that the difference between $T_{ph}$ and $T_m$ cannot be fully ignored, and thus the magnon-phonon interaction affects the magnon diffusion process, which would introduce another length scale shorter than $\lambda_m$. 

Secondly, the magnons may not follow a purely diffusive motion when they are excited. As magnons are quasi-particles, it is possible that they gain certain momentum when they are excited, for instance from the electrons in Pt. The mass of magnons at energies around $k_BT$ is roughly 1 to 2 orders of magnitude larger than the mass of electrons. In the electrical injection case, as the electrons reflect from the YIG/Pt interface, they need to transfer a vertical momentum to the magnons. This will deviate the magnon transport from a fully diffusive picture, as the magnons prefer to go vertically into the YIG film. Though this picture requires a relatively large magnon mean free path at room temperature.

Finally, as our model pertains to magnons only, we cannot fully exclude that a phononic heat-related process with an associated length scale also gives a contribution to our observed signals.

\section{V.~~~Conclusions}

We have studied the YIG thickness dependence of the nonlocal transport behavior for both the electrically and thermally excited magnons. We investigated YIG thicknesses from 0.21 $\mu$m up to 50 $\mu$m and found that the nonlocal signals of the electrically injected magnons reduce monotonically as the YIG thickness increases. Furthermore, we observed sign reversals of the nonlocal signals for the thermally injected magnons, the distance of which depends on both the heater transparency and the YIG thickness. The qualitative agreement between our results and the bulk spin Seebeck model indicates the necessity to include the magnon current and magnon accumulation in the SSE picture. Using a 2D-model we estimate the bulk spin Seebeck coefficient to be 4.5 $\mu$V/K. Our results also suggest that more complex physics processes are involved, which cannot be captured by the magnon diffusion-relaxation model. For instance, additional length scales may need to be introduced to describe the short-distance regime, or the excitation process of magnons cannot be described in a fully diffusive picture. 

\begin{figure*}
	\includegraphics[width=14cm]{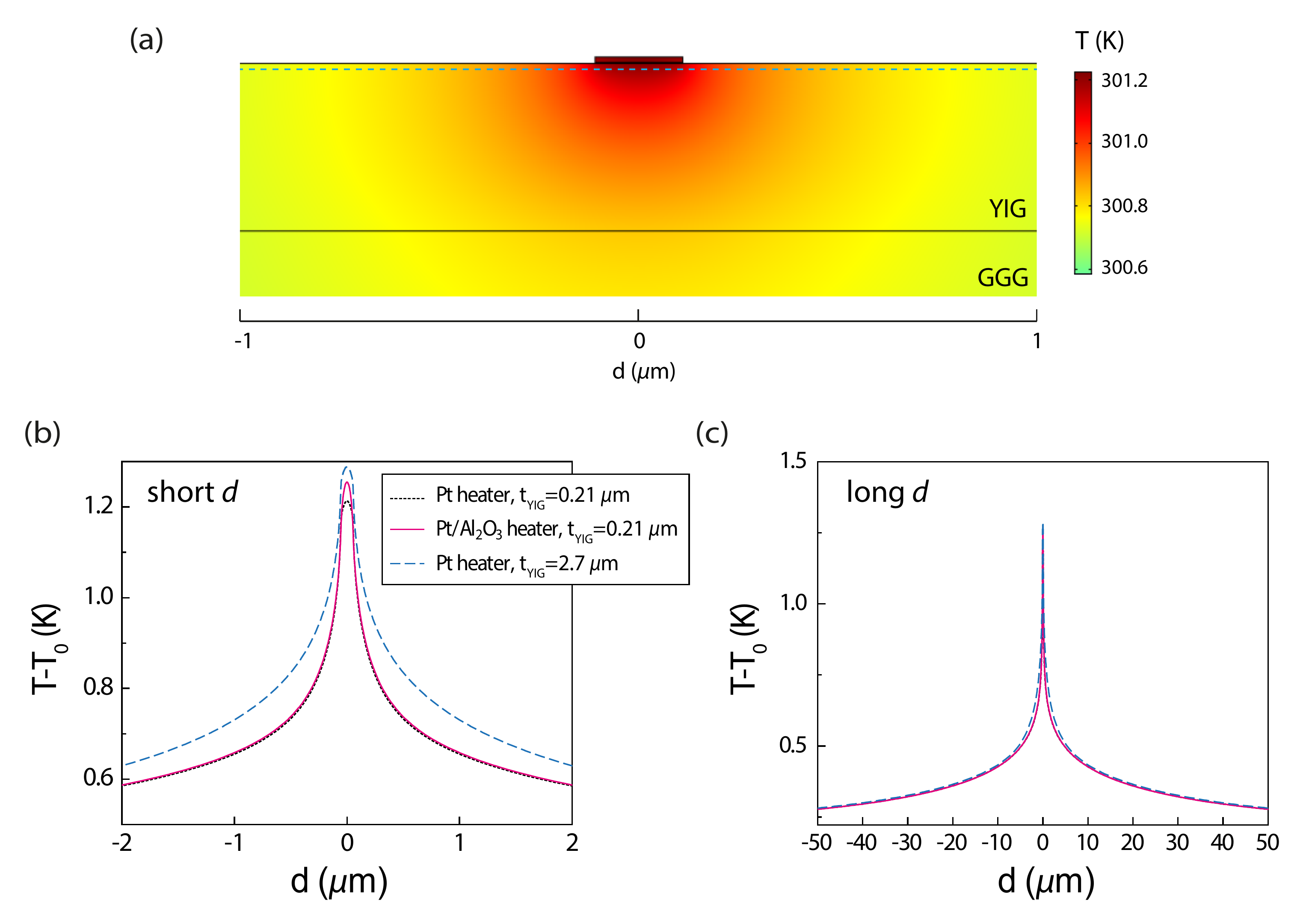}
	\caption{Temperature profile of the device induced by Joule heating. (a), Two-dimensional temperature profile close to the heat source, for Pt heater and 0.21 $\mu$m-thick YIG. The room temperature $T_0=$300 K is set as a boundary condition at the bottom of GGG. (b), (c), Temperature profiles for different heating configuration and YIG thickness along the cut line in the YIG, which is 1 nm beneath the YIG surface, as indicated by the dash line in (a). (b) shows the short-distance range and (c) shows the long-distance range.}
	\label{fig8}
\end{figure*}

\section{acknowledgments}

We thank Gerrit Bauer, Sebastian Goennenwein, Mathias Kl\"{a}ui, Yaroslav Tserkovnyak, Jiang Xiao, Yanting Chen, Jiansen Zheng and Jing Liu for inspiring discussions, M. de Roosz, H. Adema, T. Schouten and J.G. Holstein for technical assistance, and Lei Liang for helpful support and comments. This work is part of the research program of the Foundation for Fundamental Research on Matter (FOM) and is supported by NanoLab NL, EU FP7 ICT Grant InSpin 612759, NanoNextNL, the priority programme DFG Spin Caloric Transport 1538 within KU3271/1-1 and the Zernike Institute for Advanced Materials.

\section{Appendix A: ~~~Temperature profiles when Pt is served as a Joule heater}

In Fig.~\ref{fig8}, we calculated the temperature profiles of the device induced by Joule heating, to compare the temperature profiles between different heater interfaces and YIG thicknesses. For the Pt/Al$_2$O$_3$ heater scenario, an additional Al$_2$O$_3$ layer is included beneath the Pt layer in the model, with a thermal conductivity of 0.15 W/(m$\cdot$K). The calculated results from the model show that the temperature profiles with and without the Al$_2$O$_3$ layer have very little difference. We also calculated the temperature profiles for thicker YIG films, as plotted when the YIG thickness is 2.7 $\mu$m in Fig.~\ref{fig8} (b) and (c). The temperature profile is not varied more than 10$\%$ with increasing YIG thickness. Clearly, the different behaviors of the nonlocal thermal signals $V_{\textup{TG}}$ between different heater opacity or different YIG thickness cannot be attributed to the temperature profiles, but the bulk property of the magnon flow, which is sensitive to the boundary conditions. 

At further distance, the elevated temperature ($T-T_0$) by Joule heating decreases on a natural logarithmic scale as a function of $d$. Notably, compared with the exponential decay of the $V_{\textup{TG}}$ in the long-$d$ regime (see Fig.~\ref{fig5}),  the temperature decay is much slower than the $V_{\textup{TG}}$ signal decay with increasing $d$. For instance, with 10 $\mu$m further away, the temperature drops by 6$\%$ and $V_{\textup{TG}}$ drops by 66$\%$. This again strongly proves that it is the magnon accumulation instead of the temperature profile that determines the $V_{\textup{TG}}$ we measured.

Given that the present data in this paper was obtained in air, one may argue that there could be some heat carried away by air, cooling the Pt detector and giving rise to an interfacial SSE driven by the temperature difference between the Pt detector and YIG. To prove that this effect is negligible, we measured the 2.7 $\mu$m YIG sample also in vacuum, and obtained almost the same results as we measured in air. One may also argue that heat could be carried away by the Ti/Au leads, and this amount of heat is proportional to $T-T_0$ at the specific distance. If the Pt detector temperature is lowered by this effect, this could generate an additional spin Seebeck voltage which is opposite in sign compared with the local SSE signal. However, the results we obtained experimentally decrease much faster than the reduction of $T-T_0$ as a function of $d$ (see Fig.~\ref{fig5} and Fig.~\ref{fig8}(c)). Based on this fact, we conclude that these effects have no influence on the measured signals.

\section{Appendix B: ~~~Local spin Seebeck effect as a function of YIG thickness}

\begin{figure}[t]
	\includegraphics[width=8.5cm]{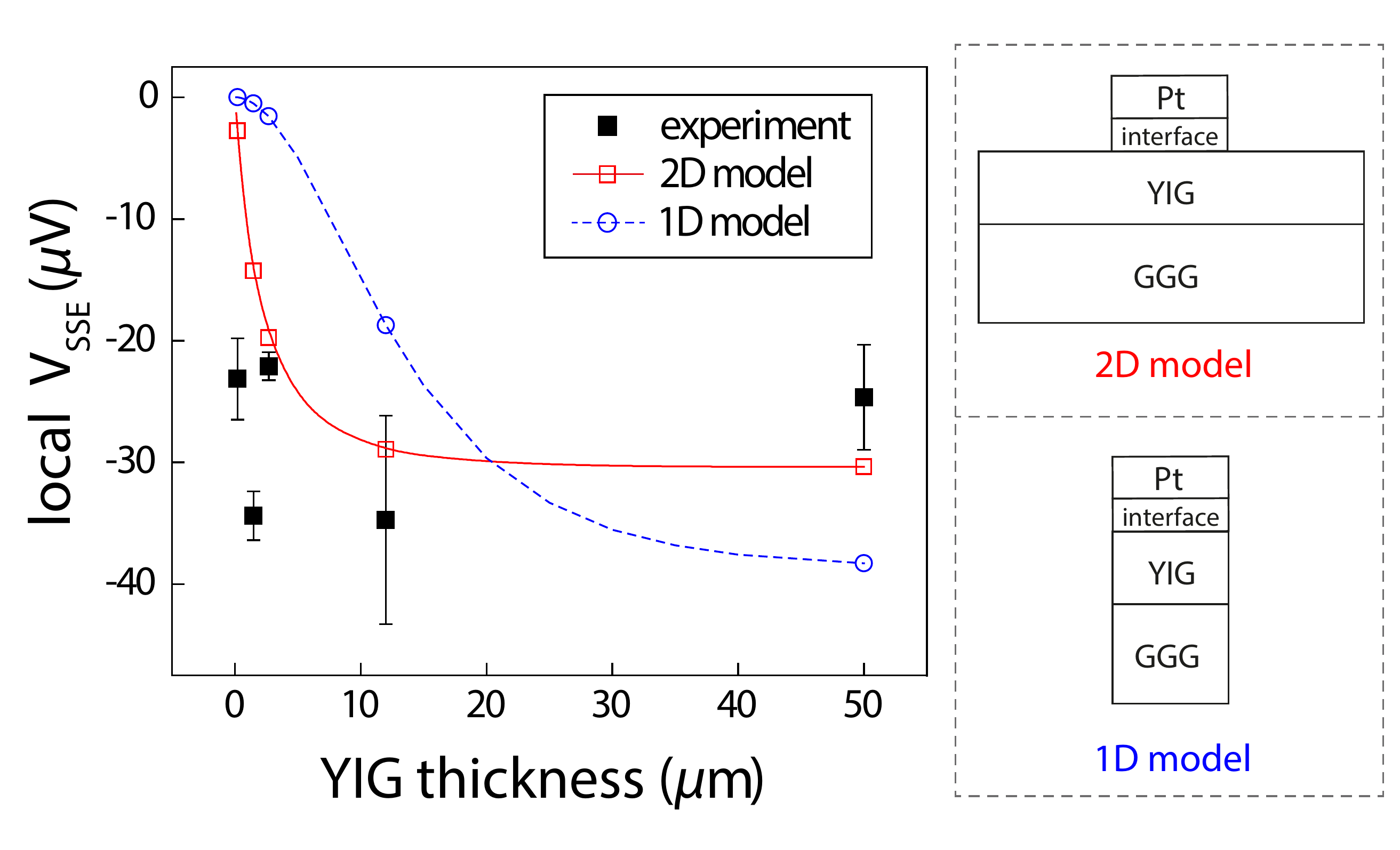}
	\caption{Local spin Seebeck voltages measured at the Pt heater strips as a function of YIG thickness. The injection current $I_{\textup{ac}}$ is 100 $\mu$A. The width and length of the Pt strips is 100 nm and 10 $\mu$m, respectively. Each point is an average from measurements over a few Pt strips and the error bars represent the standard deviations. Red curve shows the modeling results when taking $\lambda_m$=9.4 $\mu$m and $S_S$ =125 $\mu$V/K. The blue curve shows the modeling results when the YIG and GGG substrates are as wide as Pt strip, with the same $\lambda_m$ and $S_S$.}
	\label{fig9}
\end{figure}

When sending an electrical current to the Pt-only strip, the local SSE can be measured as the $V^{2f}$ signal generated at the Pt strip itself \cite{schreier_current_2013,vlietstra_simultaneous_2014}. Note that for the Pt/Al$_2$O$_3$ heater, the SSE signal vanishes, as the Al$_2$O$_3$ layer fully blocks the interaction between Pt and YIG. As shown in Appendix C and also in Ref. \cite{kehlberger_length_2015}, from the dependence of the local SSE on YIG thickness we can obtain an estimation of $\lambda_m$.

Fig.~\ref{fig9} shows the local $V_{\textup{TG}}$ results as a function of YIG thickness. It can be seen that the local $V_{\textup{TG}}$ for the different thick YIG samples are comparable. No clear trend for $V_{\textup{TG}}$ can be observed as a function of YIG thickness. This behavior clearly contradicts with the modeled results (red curve in Fig.~\ref{fig9}), using the $\lambda_m$ we extracted from the long $d$ regime from the nonlocal SSE measurements. Furthermore, the local SSE is roughly one order of magnitude larger than the largest nonlocal SSE signal we obtained, which requires a much larger $S_S$. We further modeled the situation where the YIG surface is fully covered by Pt, with the same charge current density sent in the Pt layer, creating the same amount of Joule heat as the 2D situation. Now the heat flow is not radial but vertical, normal to the plane, as shown in the blue dashed curve in Fig.~\ref{fig9}. In this case the SSE signal would saturate at larger value for YIG thickness, compared to the 2D model.

Our results suggest that the length scale that governs the local SSE can be different from the $\lambda_m$ that we extracted from the nonlocal SSE signals. As the local detection corresponds to the limit where $d \rightarrow 0$, this further confirms that for local or very short distances, more complex physics is involved.

\section{Appendix C: ~~~Vertical One-dimensional analytical model for the spin Seebeck effect}

In this section, we analytically solve a simple one-dimensional model from the bulk SSE theory \cite{rezende_magnon_2014,duine_spintronics_2015} to give a clear qualitative picture and relate it to our experimental results.

\begin{figure}
	\includegraphics[width=8.5cm]{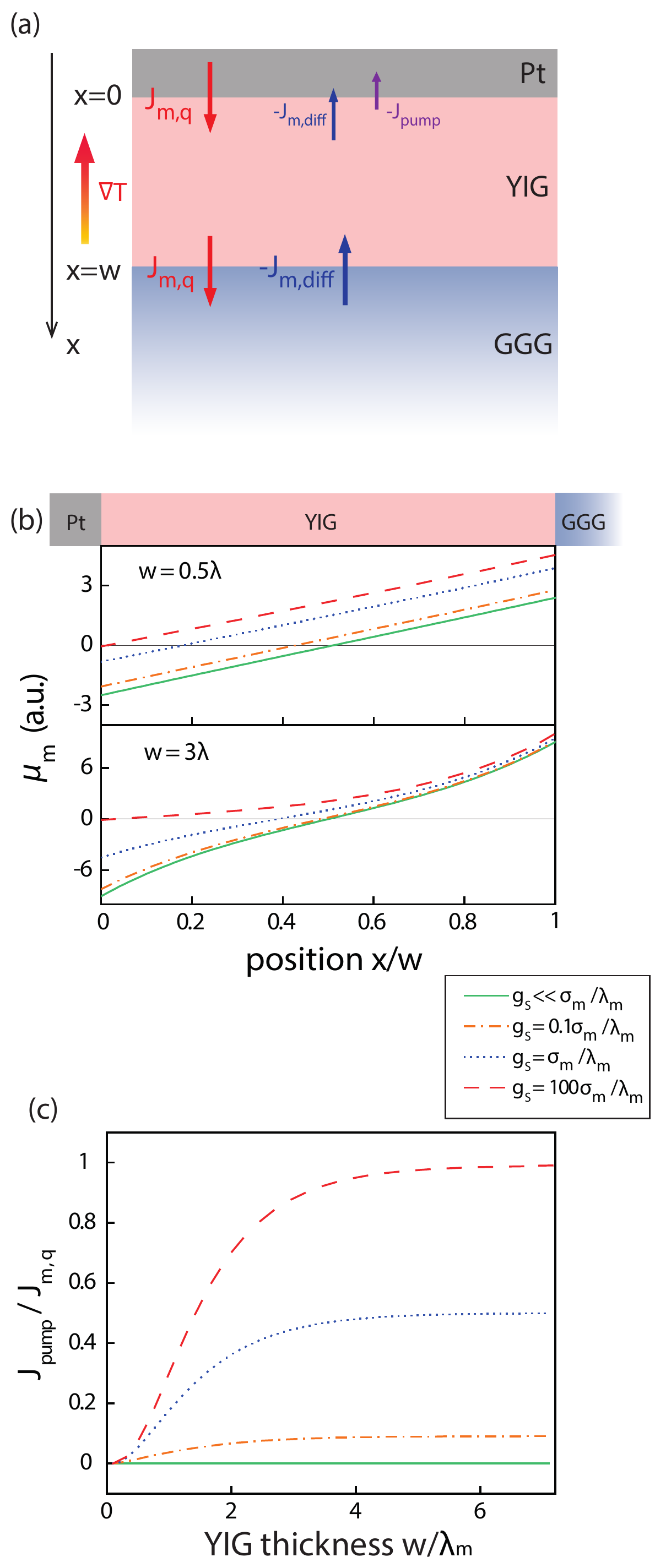}
	\caption{The application of the bulk magnonic Seebeck model to a one-dimensional vertical geometry. (a) Schematic of the Pt/YIG/GGG trilayer structure, with magnon currents only shown at the interfaces. Pt is the hotter side. (b) The calculated spatial distribution of the magnon accumulation in the YIG for different $g_S$ compared with $\sigma_m/ \lambda_m$. We take $w=0.5\lambda_m$ in the top figure and $w=3\lambda_m$ in the bottom. (c) The calculated pumping current as a function of YIG thickness for different $g_S$.}
	\label{fig10}
\end{figure}

Consider a standard triple structure where YIG is sandwiched by Pt and GGG, as shown in Fig.~\ref{fig10}(a). The heat flow, $J_q$, generated by the Joule heating in Pt, flows through the YIG uniformly towards the GGG side. From the bulk magnonic Seebeck model, a thermal magnon flow is induced in the YIG, directly proportional to $J_q$ :
\begin{equation}
J_{m,q}=- \sigma_m S_S \frac{d}{dx} T(x) \ \propto \  J_q=-\kappa \frac{d}{dx} T(x),
\label{eq:Jmq2}
\end{equation}
where $\sigma_m$ is the magnon conductivity, $S_S$ the bulk spin Seebeck coefficient and $\kappa$ the thermal conductivity of YIG, as defined in the main text. Here the temperatures of the magnon and phonon systems are assumed to be equal. On the other hand, the gradient of the magnon accumulation $\mu_m$ drives a diffusive magnon current 
\begin{equation}
J_{m,diff}=-\sigma_m\frac{d}{dx} \mu_m (x),
\label{eq:Jmdiff}
\end{equation}
where $\sigma_m$ is the magnon conductivity in YIG. From the drift-diffusion model we also have \cite{zhang_magnon_2012}:
\begin{equation}
\frac{d^2}{dx^2}\mu_m (x) =\frac{1}{\lambda_m^2}\mu_m (x),
\label{eq:diffision}
\end{equation}
where $\lambda_m$ is the magnon diffusion length of YIG. The general solution  to Eq.~\ref{eq:diffision} is
\begin{equation}
\mu_m (x) =A\exp(-\frac{x}{\lambda_m})+B\exp(\frac{x}{\lambda_m})
\label{eq:solution}
\end{equation}
with coefficients $A$ and $B$ that are determined by the boundary conditions. At $x=w$ (the YIG/GGG interface), we assume no magnon current can flow through, and therefore the total magnon current $J_m=J_{m,q}+J_{m,diff}$ should vanish to 0. At $x=0$ (the YIG/Pt interface), $J_m$ is equal to the net pumping current $J_{pump}=g_S \cdot \mu_m (0)$, where $g_S$ is the effective spin mixing conductance between YIG and Pt \cite{flipse_observation_2014,xiao_transport_2015}. These constraints set the Neumann boundary conditions for Eq.~\ref{eq:diffision}, and we can then solve $A$ and $B$ as
\begin{equation}
A=J_{m,q}\cdot\frac{1-(1-\frac{\lambda_m}{\sigma_m}g_{S})\exp(-\frac{w}{\lambda_m})}{\frac{\sigma_m}{\lambda_m}[\exp(-\frac{2w}{\lambda_m})-1]-g_{S}[\exp(-\frac{2w}{\lambda_m})+1]} \nonumber
\end{equation}
and
\begin{equation}
B=J_{m,q}\cdot\frac{\lambda_m}{\sigma_m} \exp(-\frac{w}{\lambda_m})+A\cdot \exp(-\frac{2w}{\lambda_m}),
\label{eq:AB}
\end{equation}
from which we can determine $\mu_m$ and $J_{pump}$, as shown in Figs.~\ref{fig10}(b) and (c). 

In Fig.~\ref{fig10}(b) we plot $\mu_m$ as a function of the spatial coordinate $x$. In the top figure where $w=0.5\lambda_m$, the  magnon relaxation effect is small. When the YIG/Pt interface is opaque ($g_S<<\sigma_m/ \lambda_m$), the two interfaces are symmetric for YIG. An equal amount of positive and negative $\mu_m$ builds up at the two ends of YIG, and $\mu_m$ changes sign exactly at the YIG center. As the top interface becomes more transparent, the whole $\mu_m$ shifts gradually up, as the $J_{pump}$ at the YIG/Pt interface takes away some negative magnon accumulation. The sign-reversal of the $\mu_m$ takes place closer and closer to the Pt side. In the limit where $g_S>>\sigma_m/ \lambda_m$, there will only be a very tiny negative $\mu_m$ at $x=0$. 

When $w$ is bigger than $\lambda_m$, as shown in the bottom figure, relaxation starts to enter the picture. The distribution of $\mu_m$ becomes curved, and the difference of the slope between $x=0$ and $x=w$ becomes more significant (except for the case when $g_S<<\sigma_m/ \lambda_m$), indicating a larger $J_{pump}$ compared to a smaller $w$. In Fig.~\ref{fig10}(c) we plot the $J_{pump}$ as a function of the YIG thickness for different $g_S$. It increases almost linearly for small $g_S$ and nearly quadratically for large $g_S$, and saturates when $w$ is comparable to a few times of $\lambda_m$. This result is similar to Fig. 5 in Ref.~\cite{rezende_magnon_2014}, which can be used to explain the thickness dependent SSE data from Ref.~\cite{kehlberger_length_2015}, although in Ref.~\cite{kehlberger_length_2015} they adopted a magnon temperature model to explain their data.

To test the bulk-generated SSE model, the most straightforward check is to directly probe $\mu_m$ along the YIG as a function of $x$ in an 1D-like structure. However, experimentally this is not easy to realize. It either requires a vertical $\nabla T$, and probe $\mu_m$ as a function of depth, or a fully in-plane $\nabla T$, and probe $\mu_m$ within a few $\lambda_m$ from the sample edges. Alternatively, in this experiment we adopt a nonlocal geometry where a charge current through a Pt strip (Joule heater) creates a radial thermal gradient (Fig.~\ref{fig3}(a)). Similar to the 1D situation, the temperature gradient induces a negative $\mu_m$ close to the heater and a positive $\mu_m$ far away. Due to the radial $\nabla T$ shape, the  $\mu_m$ distribution now ``goes around" and becomes detectable at the YIG surface. If we place a detector next to the heater that can sense the $\mu_m$ at the surface, it should detect negative $\mu_m$ for short distances and positive $\mu_m$ for long distances. If the YIG/heater interface is more opaque, this sign reversal should take place at a longer distance as a larger negative $\mu_m$ is preserved, same as what we observed in the experimental results.

\begin{figure}
	\includegraphics[width=8.5cm]{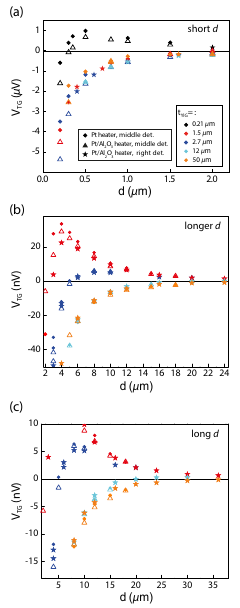}
	\caption{The linear-scale plots of the data in Fig.~\ref{fig5}. Color indicates YIG thickness and symbol shape distinguishes different heating configuration, same as defined in Fig.~\ref{fig5}.}
	\label{fig11}
\end{figure}

\section{Appendix D: ~~~Linear-scale plots of $V_{\textup{TG}}$ for different YIG thickness}

In this appendix we replot the thermally generated nonlocal signals $V_{\textup{TG}}$ for different YIG thickness and heating configurations, shown in Fig.~\ref{fig5}, all in linear scale. Note that for the longer distance plots (Fig.~\ref{fig11}(b)(c)) the $y-$axes are significantly zoomed in comparison with the full scale (Fig.~\ref{fig11}(a)), so that the sign-reversals for thicker YIG samples can be resolved. In the short-$d$ regime, except for the thin 0.21 $\mu_m$-thick YIG, all the YIG samples show similar behavior. At further distance, the sign reversals gradually take place, and moves towards a further distance for thicker YIG film.

\end{document}